\documentstyle{article}

\newcommand{\be}{\begin{equation}}
\newcommand{\ee}{\end{equation}}
\newcommand{\ben}{\begin{eqnarray}}
\newcommand{\een}{\end{eqnarray}}

\begin{document}

\title{\bf Chaplygin gas with non-adiabatic perturbations}

\author{Winfried Zimdahl\footnote{Electronic address:
zimdahl@thp.uni-koeln.de}\\
Institut f\"ur Theoretische Physik, Universit\"at zu K\"oln\\
D-50937 K\"oln, Germany\\
and\\
J\'ulio C. Fabris\footnote{Electronic address:
fabris@cce.ufes.br}\\
Departamento de F\'{\i}sica
Universidade Federal do Esp\'{\i}rito Santo\\
CEP29060-900 Vit\'oria, Esp\'{\i}rito Santo, Brazil}

\date{\today}

\maketitle

\begin{abstract}
Perturbations in a Chaplygin gas, characterized by an equation of
state $p = -A/\rho$, may acquire non-adiabatic contributions if
spatial variations of the parameter $A$ are admitted. This feature
is shown to be related to a specific internal structure of the
Chaplygin gas. We investigate how perturbations of this type
modify the adiabatic sound speed. A reduction of the effective
sound speed compared with the adiabatic value is expected to
suppress oscillations in the matter power spectrum. This text is
an abridged version of the following reference: W. Zimdahl and
J.C. Fabris, Classical and Quantum Gravity, {\bf 22}, 4311(2005).
\end{abstract}

\section{Introduction}

Since 1998 \cite{SN} a growing number of observational data has
backed up the conclusion that the expansion rate of our present
Universe is increasing. According to our current understanding on
the basis of Einstein's General Relativity, such a dynamics
requires a cosmic substratum with an effective negative pressure.
To clarify the physical nature of this substratum is one of the
major challenges in cosmology. Most of the approaches in the field
rely on a two-component picture of the cosmic medium with the (at
present) dynamically dominating "Dark Energy" (DE), equipped with
a negative pressure, which contributes roughly 70\% to the total
energy density, and with pressureless (Cold) "Dark Matter" (CDM),
which contributes roughly 30\% (see, e.g. \cite{Dark} and
references therein). Usually, these components are assumed to
evolve independently, but more general interacting models have
been considered as well \cite{IQ}.
\par
A single-component model which has attracted some interest as an
alternative description is a Chaplygin gas \cite{Chaplygin}. The
Chaplygin gas which is theoretically based in higher dimensional
theories \cite{Jackiw}, has been considered as a candidate for a
unified description of dark energy and dark matter
\cite{Kamen,bilic,Julio,bento1,Finelli1,bento2}. Its energy
density smoothly changes from that of matter at early times to an
almost constant value at late times. It interpolates between a
phase of decelerated expansion, necessary for structure formation
to occur, and a subsequent period in which the dynamically
dominating substratum acts similarly as a cosmological constant,
giving rise to accelerated expansion.
\par
The Chaplygin gas combines a negative pressure with a positive
sound velocity. This sound speed is negligible at early times and
approaches the speed of light in the late-time limit. A sound
speed of the order of the speed of light has implications which
apparently disfavor the Chaplygin gas as a useful model of the
cosmic medium. In particular, it should be connected with
oscillations of the medium on small (sub-horizon) scales. The fact
that the latter are not observed has led the authors of
\cite{Sandvik} to the conclusion that Chaplygin gas models of the
cosmic medium are ruled out as competitive candidates.
\par
However, these conclusions rely on the assumption of an adiabatic
cosmic medium. It has been argued that there might exist entropy
perturbations, so far not taken into account, which may change the
result of the adiabatic perturbation analysis \cite{NJP,Reis}. A
problem here is the origin of non-adiabatic perturbations which
should reflect the internal structure of the cosmic medium. The
latter is unknown but it may well be more complicated then
suggested by the usually applied simple (adiabatic) equations of
state. Generally, non-adiabatic perturbations will modify the
adiabatic sound speed.
\par
The purpose of this paper is to study a simple model of
non-adiabatic perturbations in a Chaplygin gas. Starting with a
two-component description of the cosmic medium we demonstrate that
any Chaplygin gas in a homogeneous and isotropic Universe can be
regarded as being composed of another Chaplygin gas which is in
interaction with a pressureless fluid such that that the
background ratio of the energy densities of both components is
constant. Spatial perturbations of this ratio then give rise to
non-adiabatic pressure perturbations which may be related to
fluctuations of the equation of state parameter $A$. The influence
of these perturbations on the sound speed is investigated.

\section{Interacting two-fluid dynamics}
\label{two-fluid dynamics}

Much of the appeal of a Chaplygin gas model of the cosmic
substratum is due to the circumstance that it represents a unified
description of dark energy and dark matter within a one component
model. In order to describe non-adiabatic features which are the
manifestation of an internal structure of the medium it will be
instructive, however, to start the discussion by establishing
basic relations of a two-fluid description. The point here is that
we shall reveal an equivalent two-component picture of any
one-component Chaplygin gas (for different two-component
decompositions see \cite{bento2,koivisto}). This circumstance can
be used to introduce a simple model for an internal structure
which gives rise to non-adiabatic pressure perturbations.

\subsection{General setting}
\label{general}

We assume the cosmic medium to behave as a perfect fluid with an
energy-momentum tensor
\begin{equation}\label{Ttot}
T^{ik} = \rho u^{i}u^{k} + p h^{ik} ,\qquad\ h^{ik} = g^{ik} +
u^{i} u^{k}\ .
\end{equation}
Our approach is based on the decomposition
\begin{equation}\label{Tsum}
T^{ik} = T_{1}^{ik} + T_{2}^{ik}
\end{equation}
of the total energy-momentum into two parts with ($A= 1, 2$)
\begin{equation}\label{TA}
T_{A}^{ik} = \rho_{A} u_A^{i} u^{k}_{A} + p_{A}^{*} h_{A}^{ik} \
,\qquad\ h_{A}^{ik} = g^{ik} + u_A^{i} u^{k}_{A} , \qquad
T_{A\,;k}^{ik} = 0\ .
\end{equation}
For our purpose it is convenient to split the (effective)
pressures $p_{1}^{*}$ and $p_{2}^{*}$ according to
\begin{equation}\label{defp*}
p_{1}^{*} \equiv p_{1} + \Pi_{1}\ ,\quad p_{2}^{*} \equiv p_{2} +
\Pi_{2}\ .
\end{equation}
This procedure allows us to introduce an arbitrary coupling
between both fluids by the requirement
\begin{equation}\label{Pi1=-Pi2}
\Pi_{1} = -\Pi_2 \equiv \Pi \quad \Leftrightarrow \quad p_{1}^{*}
= p_{1} + \Pi\ ,\quad p_{2}^{*} = p_{2} - \Pi\ .
\end{equation}
Under this condition the apparently separate energy-momentum
conservation $T_{A\,;k}^{ik} = 0$ in (\ref{TA}) is only formal.
Then, for a homogeneous and isotropic universe with $u_1^{i} =
u^{i}_{2} = u^{i}$ and $u^{a}_{;a} = 3H$, where $H$ is the Hubble
rate, the individual energy balance equations are coupled and take
the form
\begin{equation}
\dot \rho_1 + 3H\left(\rho_1 + p_1 \right) = - 3H\Pi \quad
\Leftrightarrow \quad \dot \rho_1 + 3 H \left(\rho_1 +
p_1^{*}\right)= 0 \ , \label{dotrho1}
\end{equation}
and
\begin{equation}
\dot \rho_2 + 3 H \left(\rho_2 + p_2 \right) = 3H\Pi \quad
\Leftrightarrow \quad \dot \rho_2 + 3 H \left(\rho_2 +
p_2^{*}\right)= 0 \ , \label{dotrho2}
\end{equation}
where $\rho = \rho_1 + \rho_2$ and $p = p_1 + p_2$. The total
adiabatic sound velocity may be split according to
\begin{equation}
\frac{\dot{p}}{\dot{\rho}} =
\frac{\dot{\rho}_{1}}{\dot{\rho}}\frac{\dot p _1}{\dot \rho _1} +
\frac{\dot{\rho}_{2}}{\dot{\rho}} \frac{\dot p _2}{\dot \rho _2} =
\frac{\dot{\rho}_{1}}{\dot{\rho}}\frac{\dot p^{*}_1}{\dot \rho _1}
+ \frac{\dot{\rho}_{2}}{\dot{\rho}} \frac{\dot p^{*}_2}{\dot \rho
_2} \ . \label{soundspeedsplit}
\end{equation}
We emphasize that so far the equations of state and the nature of
the interaction between the components are left unspecified.

\subsection{Matter perturbations}
\label{matter perturbations}

To study perturbations about the homogeneous and isotropic
background we introduce the following quantities. We define the
total fractional energy density perturbation
\begin{equation}
D \equiv \frac{\hat{\rho}}{\rho+ p} = - 3H
\frac{\hat{\rho}}{\dot\rho}\ , \label{D}
\end{equation}
and the energy density perturbations for the components ($A=1,\,
2$)
\begin{equation}
D_{A} \equiv \frac{\hat{\rho}_{A}}{\rho _{A}+ p_{A}^{*}} = - 3H
\frac{\hat{\rho}_A}{\dot\rho_A} \ .\label{DA}
\end{equation}
Likewise, the corresponding pressure perturbations are
\begin{equation}
P \equiv \frac{\hat{p}}{\rho+ p} = - 3H \frac{\hat{p}}{\dot\rho} =
- 3H \frac{\dot{p}}{\dot\rho}\frac{\hat{p}}{\dot p}\ ,
\label{Ptot}
\end{equation}
and
\begin{equation}
P_{A} \equiv \frac{\hat{p}_{A}}{\rho _{A}+ p_{A}^{*}} = - 3H
\frac{\hat{p}_A}{\dot\rho_A} = - 3H
\frac{\dot{p}_{A}}{\dot{\rho}_A}\frac{\hat{p}_A}{\dot p_A}\ .
\label{PA}
\end{equation}
One also realizes that
\begin{equation}
D = \frac{\dot\rho_{1}}{\dot\rho} D_{1}+
\frac{\dot\rho_{2}}{\dot\rho}D_{2} \ \label{Dsum}
\end{equation}
and
\begin{equation}
P = \frac{\dot\rho_{1}}{\dot\rho} P_{1} +
\frac{\dot\rho_{2}}{\dot\rho} P_{2} \ .\label{Psum}
\end{equation}
In terms of these quantities the non-adiabatic pressure
perturbations
\begin{equation}
P - \frac{\dot{p}}{\dot{\rho}}D = - 3H\frac{\dot{p}}{\dot\rho}
\left[\frac{\hat{p}}{\dot p} - \frac{\hat{\rho}}{\dot\rho}\right]
\ \label{defPnad}
\end{equation}
are then generally characterized by
\begin{eqnarray}
P - \frac{\dot{p}}{\dot{\rho}}D &=& \frac{\dot\rho_{1}}{\dot\rho}
\left(P_1 - \frac{\dot{p}_1}{\dot{\rho}_1}D_1 \right) +
\frac{\dot\rho_{2}}{\dot\rho}
\left(P_2 - \frac{\dot{p}_2}{\dot{\rho}_2}D_2 \right)\nonumber\\
&& + \frac{\dot\rho_{1}\dot\rho_{2}}{\left(\dot\rho \right)^2}
\left[\frac{\dot{p}_2}{\dot{\rho}_2} -
\frac{\dot{p}_1}{\dot{\rho}_1} \right] \left[D_2 - D_1\right]\ .
\label{Pnad}
\end{eqnarray}
\ \\
The first two terms on the right-hand side describe internal
non-adiabatic perturbations within the individual components. The
last term takes into account non-adiabatic perturbations due to
the two-component nature of the medium.

\section{The Chaplygin gas}
\label{Chaplygin}

\subsection{Two-component interpretation}
\label{two-component}

The Chaplygin gas is characterized by an equation of state

\begin{equation}
p = - \frac{A}{\rho}\ , \label{eosChap}
\end{equation}
where $p$ is the pressure and $\rho$ is the energy density of the
gas. $A$ is a positive constant. This equation of state gives rise
to the energy density \cite{Kamen}
\begin{equation} \rho = \sqrt{A + \frac{B}{a^6}}\ ,
\label{rhoChap}
\end{equation}
where $B$ is another (positive) constant. Now, let us rename the
constants according to
\begin{equation}
A = A_2\left(1+\kappa\right)\ , \qquad B =
B_2\left(1+\kappa\right)^2\ . \label{defA2}
\end{equation}
For any constant, non-negative $\kappa$ the quantities $A_2$ and
$B_2$ are constant and non-negative as well. So far, no physical
meaning is associated with these constants. It is obvious, that
this split allows us to write the energy density (\ref{rhoChap})
as
\begin{equation}
\rho = \left(1+\kappa\right)\sqrt{\frac{A_2}{1+ \kappa} +
\frac{B_2}{a^6}}\ . \label{rhoviaA2}
\end{equation}
This means, $\rho$ can be regarded as consisting of two
components, $\rho_{1}$ and $\rho_{2}$,
\begin{equation}
\rho = \rho_1 + \rho_2\ , \label{5}
\end{equation}
with
\begin{equation}
\rho_1 = \sqrt{\frac{A_2\kappa^2}{1+ \kappa} +
\frac{B_2\kappa^2}{a^6}}\ \label{rho1} \ee and \be \rho_2 =
\sqrt{\frac{A_2}{1+ \kappa} + \frac{B_2}{a^6}} \ , \label{rho2}
\end{equation}
where $\rho_1 = \kappa\rho_2$ is valid, i.e., the ratio
$\rho_1/\rho_2 = \kappa$ is constant. The corresponding
(effective) equations of state are (the notations are chosen in
agreement with the formalism of subsection \ref{general})
\begin{equation}
p_1^{*} = - \frac{A_2 \kappa^2}{\left(1+\kappa\right)\rho_1} = -
\frac{A_2 \kappa}{\left(1+\kappa\right)\rho_2} \ , \label{p1star}
\end{equation}
and
\begin{equation}
p_2^{*} = - \frac{A_2}{\left(1+\kappa\right)\rho_2} \ .
\label{p2star}
\end{equation}
This implies $p_1^{*} = \kappa p_2^{*}$, i.e., the effective
pressures of both components differ by the same constant which
also characterizes the ratio of both energy densities. It is
further convenient to introduce a quantity
\begin{equation}
p_2 \equiv - \frac{A_2}{\rho_2}\ . \label{p2}
\end{equation}
This quantity differs from $p_2^{*}$ by
\begin{equation}
p_2^{*} - p_2 = \frac{\kappa A_2}{\left(1+\kappa\right)\rho_2} = -
p_1^{*}\ . \label{12}
\end{equation}
Defining also a quantity $\Pi$ (the use of the same symbol as in
(\ref{Pi1=-Pi2}) will be justified in the next step)  by
\begin{equation}
\Pi \equiv p_1^{*} = - \frac{\kappa}{1+\kappa}\frac{A_2}{\rho_2} =
- \frac{\kappa}{1+\kappa}\frac{A}{\rho}\ , \label{Pip1star}
\end{equation}
one checks by direct calculation that the following relations are
valid:
\begin{equation}
\dot \rho_1 + 3 H \rho_1 = - 3H\Pi \quad \Leftrightarrow \quad
\dot \rho_1 + 3 H \left(\rho_1 + p_1^{*}\right)= 0 \ ,
\label{dotrho1n}
\end{equation}
and
\begin{equation}
\dot \rho_2 + 3 H \left(\rho_2 + p_2 \right) = 3H\Pi \quad
\Leftrightarrow \quad \dot \rho_2 + 3 H \left(\rho_2 +
p_2^{*}\right)= 0 \ . \label{dotrho2n}
\end{equation}

At this point it becomes clear what we have obtained by the formal
manipulations in Eq.(\ref{defA2}). It describes a two-component
system in which the components interact with each other. The role
of the interaction is to keep the ratio $\kappa$ of the energy
densities of both components constant. In other words, {\it any
given Chaplygin gas can be thought as being composed of another
Chaplygin gas which is in interaction with a pressureless fluid
such that the energy density ratio of both components is fixed.}

The components have sound velocities which are different from each
other and different from the overall sound velocity. For the
latter we have
\begin{equation}
\frac{\dot p}{\dot \rho} = - \frac{p}{\rho} = \frac{A}{\rho^2} =
\frac{A}{A + \frac{B}{a^6}} \ . \label{soundChap}
\end{equation}
The sound velocities of the components are $\dot p _1 /\dot \rho
_1 = 0$ and
\begin{equation}
\frac{\dot p _2}{\dot \rho _2} = - \frac{p_2}{\rho_2} =
\frac{A_2}{\rho_2^2} = \frac{A_2}{\frac{A_2}{1 + \kappa} +
\frac{B_2}{a^6}} = \left(1 + \kappa\right)\frac{\dot p}{\dot \rho}
\ . \label{sound2}
\end{equation}
Since for large times we have $\dot{p}/\dot{\rho} \rightarrow 1$,
the quantity $\dot{p}_{2}/\dot{\rho}_2$ may be larger than unity
which at the first glance seems to imply a superluminal sound
propagation. However, $\dot{p}_{2}/\dot{\rho}_2$ is a formal
quantity only, which does not describe any propagation phenomenon.
The adiabatic sound speed (\ref{soundChap}) may also be split with
respect to the effective presures according to
(\ref{soundspeedsplit}) with
\begin{equation}
\frac{\dot p^{*}_1}{\dot \rho _1} =
\frac{\dot{p}}{\dot{\rho}}\qquad {\rm and} \qquad \frac{\dot
p^{*}_2}{\dot \rho _2} = \frac{\dot{p}}{\dot{\rho}} \ .
\label{23c}
\end{equation}
These effective sound velocities of the components coincide with
the total sound velocity.

\section{Non-adiabatic pressure perturbations}
\label{non-adiabatic}

The Chaplygin in its two-component interpretation of the previous
section belongs to a class of models with
\begin{equation}
\dot{\kappa} = 0 \quad {\rm and}\quad \rho_1 + p_1^{*} = \kappa
\left(\rho_2 + p_2^{*}\right)\ . \label{constkappa}
\end{equation}
Generally, the components may have (not necessarily constant)
equations of state
\\
\begin{equation}
p_{1} = w_{1} \rho_{1}\quad{\rm and}\quad p_{2} = w_{2} \rho_{2}\
. \label{}
\end{equation}
For the total equation of state parameter $w$ it follows that
\begin{equation}
w = \frac{w_2 + \kappa w_1}{1 + \kappa}\quad{\rm and}\quad p = w
\rho \ . \label{}
\end{equation}
For our special Chaplygin gas case these quantities are
\begin{equation}
w_1 = 0 ,\quad w_2 = - \frac{A_2}{\rho_{2}^{2}} ,\quad w = -
\frac{A}{\rho^{2}} \ . \label{w}
\end{equation}
Under the condition (\ref{constkappa}) the difference of the
individual sound speeds is
\begin{equation}
\frac{\dot{p}_2}{\dot{\rho}_2} - \frac{\dot{p}_1}{\dot{\rho}_1} =
\frac{1 + \kappa}{\kappa} \frac{\dot{\Pi}}{\dot{\rho}_2} \ .
\label{}
\end{equation}
Introducing now matter perturbations in terms of the quantities
defined in subsection \ref{matter perturbations} and allowing the
density ratio $\kappa$ to fluctuate, the difference between the
fractional energy density perturbations which describes entropy
perturbations becomes directly proportional to $\hat \kappa$:
\begin{equation}
D_2 - D_1 =  - \frac{1}{1 + w}\,\frac{\hat \kappa}{\kappa} \ .
\label{D2-D1}
\end{equation}
This demonstrates, that a perturbation of the density ratio is
essential for entropy perturbations to occur. Since the background
ratio $\kappa$ also determines the relation between the equation
of state parameters $A$ and $A_2$ in (\ref{defA2}) a further
specification may be performed. We shall assume from now on the
parameter $A_2$ to be a true constant, i.e., $\hat{A}_2 =0$. This
choice implies
\begin{equation}
\frac{\hat{A}}{A} = \frac{\hat{\kappa}}{1 + \kappa} \ .
\label{hatAChap}
\end{equation}
Within this model a fluctuation of the density ratio is equivalent
to a fluctuation of the equation of state parameter $A$. Any
fluctuation of $\kappa$ or $A$ generates a non-adiabatic pressure
fluctuation
\begin{equation}
P - \frac{\dot{p}}{\dot{\rho}}D =
\frac{w}{1+w}\,\frac{\hat{\kappa}}{1 + \kappa} =
\frac{w}{1+w}\,\frac{\hat{A}}{A}\ . \label{PnaChap}
\end{equation}
We conclude that perturbing the equation of state parameter $A$
represents a way to introduce non-adiabatic pressure perturbations
in a Chaplygin gas. It is expedient to recall that in an
underlying string theoretical formalism the parameter $A$ is
connected with the interaction strength of d-branes \cite{Jackiw}.
Hence, a fluctuating $A$ corresponds to fluctuating interactions
in string theory. An alternative, perhaps more transparent way of
understanding the meaning of the fluctuation of the parameter $A$
is to consider the Born-Infeld action that leads to the Chaplygin
gas. The Lagrangian density in this case takes the form
\cite{sen,copeland}
\begin{equation}
L = \sqrt{-g}V(\phi)\sqrt{- det[g_{\mu\nu} +
\phi_{;\mu}\phi_{;\nu}]} \quad , \label{LBI}
\end{equation}
where $V(\phi)$ is a potential term. This Lagrangian density leads
to the Chaplygin gas equation of state for \cite{bento1,bento2}
\begin{equation}
V(\phi) = \sqrt{A}\ . \label{VA}
\end{equation}
For $\phi = \phi_0 + \hat{\phi}$, the fluctuation in $A$ is given
by
\begin{equation}
\frac{\hat{A}}{A}= 2 \left(\frac{V'}{V}\right)_{\phi =
\phi_0}\hat{\phi} \quad ,
\end{equation}
where the prime denotes the derivative with respect to $\phi$.
Consequently, fluctuations of $A$ are allowed if the scalar field
$\phi$ is not in the minimum of the potential. It has been shown
that there are configurations for which a sufficiently flat
potential (equivalent to an almost constant $A$) admits
accelerated expansion \cite{copeland}.

Given that $w<0$, any $\hat{A}
> 0$ will reduce the adiabatic pressure perturbations.
For definite statements further assumptions about $\hat{A}$ (or
$\hat{\kappa}$) are necessary since otherwise the problem is
undetermined. The structure of (\ref{PnaChap}) motivates a choice
$\hat{A}/A = \mu\left(1 + w\right) D$, for which
\begin{equation}
P = c_{eff}^{2} D \ ,\quad c_{eff}^{2} \equiv
\frac{\dot{p}}{\dot{\rho}}\left(1 - \mu\right)\ . \label{Pceff}
\end{equation}
Any $\mu$ in the range $0<\mu\leq 1$ leads to an effective sound
speed square $c_{eff}^{2}$ which is reduced compared to the
adiabatic value $ \dot{p}/\dot{\rho} = -w$. It is interesting to
note that the relation
\[
\hat{p} = \left(1 - \mu\right)\frac{\dot{p}}{\dot{\rho}}\hat{\rho}
= - \left(1 - \mu\right)\frac{p}{\rho}\hat{\rho} \ ,
\]
which is a different way of writing (\ref{Pceff}), coincides with
that of a generalized Chaplygin gas with an equation of state $p =
- C/\rho^{\left(1 - \mu\right)}$ with (a true) constant $C$. For a
generalized Chaplygin gas one would, however, also have $\dot{p} =
- \left(1 - \mu\right) \frac{p}{\rho}\dot{\rho}$ and hence
adiabatic perturbations only. In a sense, our strategy to consider
perturbations $\hat{A}\neq 0$ while $\dot{A} =0$, implies that the
medium behaves as a Chaplygin gas in the background and shares
features of a generalized Chaplygin gas on the perturbative level.

At this point also the role of the assumption $\hat{A}_2 =0$ that
precedes eq.~(\ref{hatAChap}) becomes clear. We mentioned at the
end of subsection \ref{two-component} that the decomposition of a
given Chaplygin gas into another Chaplygin gas interacting with
matter may be repeated again and again. In a subsequent step $A_2$
would play the role that $A$ played so far. Hence, the assumption
$\hat{A}_2 =0$ ensures that no further non-adiabatic contributions
will appear.

\section{Conclusions}
\label{conclusions}

We have demonstrated that any Chaplygin gas in a homogeneous and
isotropic Universe can be regarded as being composed of another
Chaplygin gas which is in interaction with a pressureless fluid
such that that the background ratio of the energy densities of
both components remains constant. This two-component
interpretation allowed us to establish a simple model for
non-adiabatic pressure perturbations as the result of a specific
internal structure of the substratum. Within the one-component
picture this internal structure manifests itself as a spatial
fluctuation of the parameter $A$ in the Chaplygin gas equation of
state $p = - A/\rho$. Such fluctuations, which can be traced back
to fluctuations of a tachyon field potential, modify the adiabatic
sound speed of the medium which may shed new light on the status
of Chaplygin gas models of the cosmic substratum.
\ \\
%\acknowledgements
{\bf Acknowledgements}

We thank the organizers of the {\it I Brazilian Workshop on Dark
Energy/Matter}, held in Joinville, Brazil, for the hospitality and for the nice ambiance
during the meeting. This work has been partially supported by CNPq (Brazil).
\ \\

\end{document}